# Wind Farm Icing Loss Forecast Pertinent to Winter Extremes

Linyue Gao [1,2], Teja Dasari [3], Jiarong Hong [1, 4, *]

[1] St. Anthony Falls Laboratory, University of Minnesota, Minneapolis, MN 55414, USA

[2] California State University, Sacramento, Sacramento, CA 95819, USA

[3] Xcel Energy, Minneapolis, MN 55401, USA

[4] Department of Mechanical Engineering, University of Minnesota, Minneapolis, MN 55455, USA

[*] Corresponding author: jhong@umn.edu (J. Hong).

**Abstract**

The 2021 Texas power crisis has highlighted the vulnerability of the power system under wind extremes, particularly with the increasing penetration of energy resources that depend on weather conditions (e.g., wind energy). The current wind power forecast models do not effectively consider the impact of such extreme weather events. In the present study, we provide a fast and robust statistical model developed using ten years of utility-scale turbine data at the Eolos Wind Energy Research Station to forecast the icing losses under such weather conditions. This model covers different cold climate impacts, including precipitation icing, frost contamination, and low-temperature effect. This model has been assessed using three large-scale (>100 MW) wind farm data involving turbines with different capacities and from different manufacturers across multiple geographic regions. Notably, the model has been used to predict the wind power losses in the entire Texas (> 91% of total wind installation in Texas) during 2021 Texas power crisis. The proposed model can be easily integrated into the existing wind farm and power grid operations, allowing the power system operators can develop more appropriate and pinpointed plans to balance the severe and sudden energy deficits and increase the system integrity under winter extremes.

## Nomenclature

| | |
|---|---|
| $a_1$ | A coefficient relating to $C_1$ |
| $a_2$ | A coefficient relating to $C_2$ |
| $a_3$ | A coefficient relating to $C_3$ |
| $a_4$ | A coefficient relating to $C_5$ |
| $a_5$ | A coefficient relating to $C_6$ |
| $c_1, \ldots, c_9$ | Regression constants for the polynomial fit of the power curve |
| $C_1$ | An incubation factor that describes the delay of the onset of WTI event in comparison to the onset of the CC period [h: hour] |
| $C_2$ | A persistence/ablation factor that corrects the delay between the end of CC period and the end of WTI event |
| $C_3$ | Correction factor relating to the ratio of the duration of operational-icing phase ($\widehat{D}_{\mathrm{OP}}$) to the total duration of a WTI event |
| $C_4$ | A scaling factor for different turbines |
| $C_5$ | A correction factor relating to wind speed |
| $C_6$ | Another correction factor relating to wind speed |
| $\widehat{D}_{\mathrm{WTI}}$ | Modeled total duration of a potential WTI event [h] |
| $\widehat{D}_{\mathrm{OP}}$ | Modeled duration of operational-icing phase of a potential WTI event [h] |
| $\widehat{D}_{\mathrm{ST}}$ | Modeled duration of stopped-icing phase of a potential WTI event [h] |
| $\widehat{D}_{\mathrm{PO}}$ | Modeled duration of post-icing phase of a potential WTI event [h] |
| $\widehat{\Delta E}_{\mathrm{WTI}}$ | Modeled energy losses in the entire WTI event [kWh] |
| $\widehat{\Delta E}_{\mathrm{WTI,WF}}$ | Modeled total energy loss of each detected WTI event in a wind farm [kWh] |
| $\widehat{\Delta E}_{\mathrm{WF}}$ | Modeled total energy loss of all WTI events of a wind farm in the forecasting period [kWh] |
| $\Delta E_{\mathrm{WF}}$ | Measured total energy loss of all WTI events of a wind farm in the forecasting period [kWh] |
| $\widehat{\Delta E}_{\mathrm{OP}}$ | Modeled energy losses in operational-icing phase [kWh] |
| $\widehat{\Delta E}_{\mathrm{PO}}$ | Modeled energy losses in post-icing phase [kWh] |
| $\widehat{\Delta E}_{ST}$ | Modeled energy losses in stopped-icing phase [kWh] |
| $i$ | $i = 1, \ldots, m$ indicates the number of the WTI events in the forecasting period |
| $n$ | Number of turbine in a wind farm |
| $P_{\max}$ | Turbine maximum capacity [kW] |
| $P_{\max,2.5\mathrm{MW}}$ | Turbine maximum capacity of 2.5MW Eolos turbine [kW] |
| $P_{\mathrm{mea}}$ | Measured turbine power [kW] |
| $P_{\mathrm{no-icing}}$ | Modeled power under no-icing conditions based on measured wind speed by the turbine nacelle anemometer [kW] |
| $\tilde{P}_{\mathrm{no-icing}}$ | Modeled power under no-icing conditions based on the wind speed from NWP [kW] |
| $R$ | Reduction factor in the model |
| $R_{\mathrm{OP}}$ | A descending sequence from 1 to 0 with a fixed step of $\Delta t/\widehat{D}_{\mathrm{OP}}$ |
| $R_{\mathrm{PO}}$ | An ascending sequence from 0 to 1 with a fixed step of $\Delta t/\widehat{D}_{\mathrm{PO}}$ |
| $RH$ | Relative humidity [%] |
| $T$ | Air temperature [ ℃] |
| $t_{\mathrm{CC,start}}$ | Start time of CC period |
| $t_{\mathrm{CC,end}}$ | End time of CC period |
| $\hat{t}_{\mathrm{WTI,start}}$ | Modeled start time of a potential WTI event |
| $\hat{t}_{\mathrm{WTI,end}}$ | Modeled end time of a potential WTI event |
| $\Delta t$ | Time interval of the input dataset [h] |
| $\Delta t_{\mathrm{abl}}$ | Ablation time [h] |
| $t_{\mathrm{rec}}$ | The first time that the air temperature rises above 0 ℃ |
| $V$ | Wind speed [m/s] |
| $V_{\mathrm{cut-in}}$ | Cut-in wind speed [m/s] |
| $V_{\mathrm{cut-out}}$ | Cut-out wind speed [m/s] |
| $V_{\mathrm{rated}}$ | Rated wind speed [m/s] |
| $\bar{V}_{\mathrm{CC}}$ | Averaged wind speed during the CC period [m/s] |
| $\bar{V}_{\mathrm{WTI}}$ | Averaged wind speed during WTI event [m/s] |
| $\bar{V}_{\mathrm{OP}}$ | Averaged wind speed in the operational-icing phase [m/s] |
| CC | Cold climate |
| CFD | Computational fluid dynamics |
| ERCOT | Electric Reliability Council of Texas |
| IPS | Ice protection system |
| ML | Machine learning |
| NWP | Numerical weather prediction |
| RMAE | Relative mean absolute error |
| SCADA | Supervisory control and data acquisition |
| WTI | Wind turbine icing |

## 1. Introduction

Renewable energy sources are considered as the most promising alternative energy sources due to their clean and uninterrupted nature. The share of renewables in total global capacity expansion reached a new historical peak of 82% in 2020, contributing to a cumulative capacity of 2,799 GW by the end of the year [1,2]. Amongst all renewables, wind energy continues to dominate such expansion, accounting for 43% (111 GW) of all new renewable additions (261 GW) [2]. Up to now, wind energy has already developed into one of the most important contributors to the modern electric grids with a cumulative capacity of 743 GW [1]. However, the growing share of electricity production from wind energy increases the stochastic nature of the power system [3]. Specifically, in contrast to conventional generation systems, such as fossil fuel power plants that can provide stable power production, wind generation yields fluctuations across different time scales associated with the seasonal, daily, and intra-minute turbulence level variations in wind resources. Under such circumstances, accurate wind power forecasts are of vital importance to achieve appropriate dispatching, commit production-consumption balance, and eventually maintain the integrity of power system, particularly for the systems with high penetrations of wind generation [4].

Researchers have been studying wind power forecasting models for decades, and such models can be categorized into physical [5–7] and statistical approaches [8–14]. Physical methods refine the numerical weather predictions (NWPs) from a certain grid point to wind turbine levels based on the detailed physical characteristics of the terrain (i.e., orography, roughness, and obstacles) and the atmospheric stability via sub-models, such as computational fluid dynamics (CFD) tools. Statistical models rely on relationships between the historical values of turbine power and the NWP variables such as wind speed, wind direction, temperature, and pressure. The state-of-the-art statistical models are mainly based on various machine learning (ML) algorithms. However, in most previous studies, no matter using physical or statistical approaches, the forecasts are only suitable for the "standard" markets that the wind power outputs are not affected by extreme weather conditions, including snowstorms, sandstorms, tornados, etc. Specifically, such models physically lose sight of the effects of such extreme weather on turbine power production. In addition, most of those statistical models trained by ML algorithms based on the historical data artificially remove such outlier data affected by extreme conditions via a data preprocessing (i.e., data clean) process to increase the effectiveness of ML models [15,16].

Unlike sandstorms or tornados with low occurrence frequencies, 20%-30% of wind installations are located in cold climate (CC) regions, suggesting one of the largest "non-standard" markets in wind energy today, according to the survey conducted by VTT Technical Research Centre [17]. Wind turbines in CC regions tend to experience frequent atmospheric icing or periods with temperatures below the operational limits of standard IEC 61400-1 ed3 wind turbines. Up to now, most wind turbines in CC regions are not equipped with ice protection systems (IPSs) and yield, to various extents, energy losses in winter. Based on the icing severity, wind turbine icing (WTI) events can be categorized into light WTI events and severe WTI events. Light WTI events refer to the situations in which wind turbines have detectable power losses (>15%) for short periods and can recover to their regular operations without stops. Severe WTI events correspond to the conditions that wind turbines stop and generate no power due to the remarkable blade ice accretion. The latter category has a far worse impact on the turbine energy production and even the integrity of power grids, especially for small independent systems, such as the Texas power system, isolated from the rest of the country and experienced blackouts due to snowstorms.

Therefore, wind power forecasts that incorporate accurate estimates of such CC impacts are highly desired in managing such vulnerabilities and providing an optimum operation for power systems. However, the current power forecast models mentioned before directly omit the icing issues. The existing icing forecast models [18–20] can only identify the occurrence of icing events without providing detailed information on potential power losses. A rare number of studies published in recent three years [21–24] enable the forecasts of icing-related power losses based on icing accumulation rate for a simplified rotating rod derived from NWP variables. The best-reported deterministic model can achieve average forecast errors of 0.14~0.25 MW per turbine for four stations in winter periods, which reduces about 50% of errors in the power forecasts without involving icing impacts [21]. Nevertheless, it should be noted that such models tend to underestimate the power losses due to a lack of consideration of the turbine responses to the icing (e.g., shut down), particularly for the severe WTI events. Moreover, such models intend to neglect the effect of the duration of a WTI event on the accumulative energy losses, which is reported as the most critical factor influencing energy losses [22]. Those limitations suggest such models fail to meet the demand of preparing the power systems for winter extremes.

Consequently, in the present study, we focus on developing a statistical model that enables fast turbine power loss forecasts via NWP variables for severe WTI events. We derive the statistical model based on the Eolos database (i.e., Eolos Wind Energy Research Station at the University of Minnesota, referred to as Eolos database hereafter) with a 2.5 MW wind turbine. Specifically, we find the empirical formulas to link the meteorological icing variables to the durations of WTI events and the corresponding energy losses. To fully consider the severe impact of the turbine shut down period on the power losses, we further partition a WTI event into three icing phases, i.e., operational-icing, stopped-icing, and post-icing phases, and derive the semi-empirical formulation of their durations and energy losses. The detailed information of the proposed statistical model is described in Section 2. This model covers three different impacts of CC, including precipitation icing, frost condensation, and low-temperature effects. Section 3 and Section 4 present the performance assessments of the proposed method at wind farm scale and power system scale, respectively. Section 5 briefly concludes the main contributions of the present study.

## 2. Methodology

### 2.1 Icing loss calculation

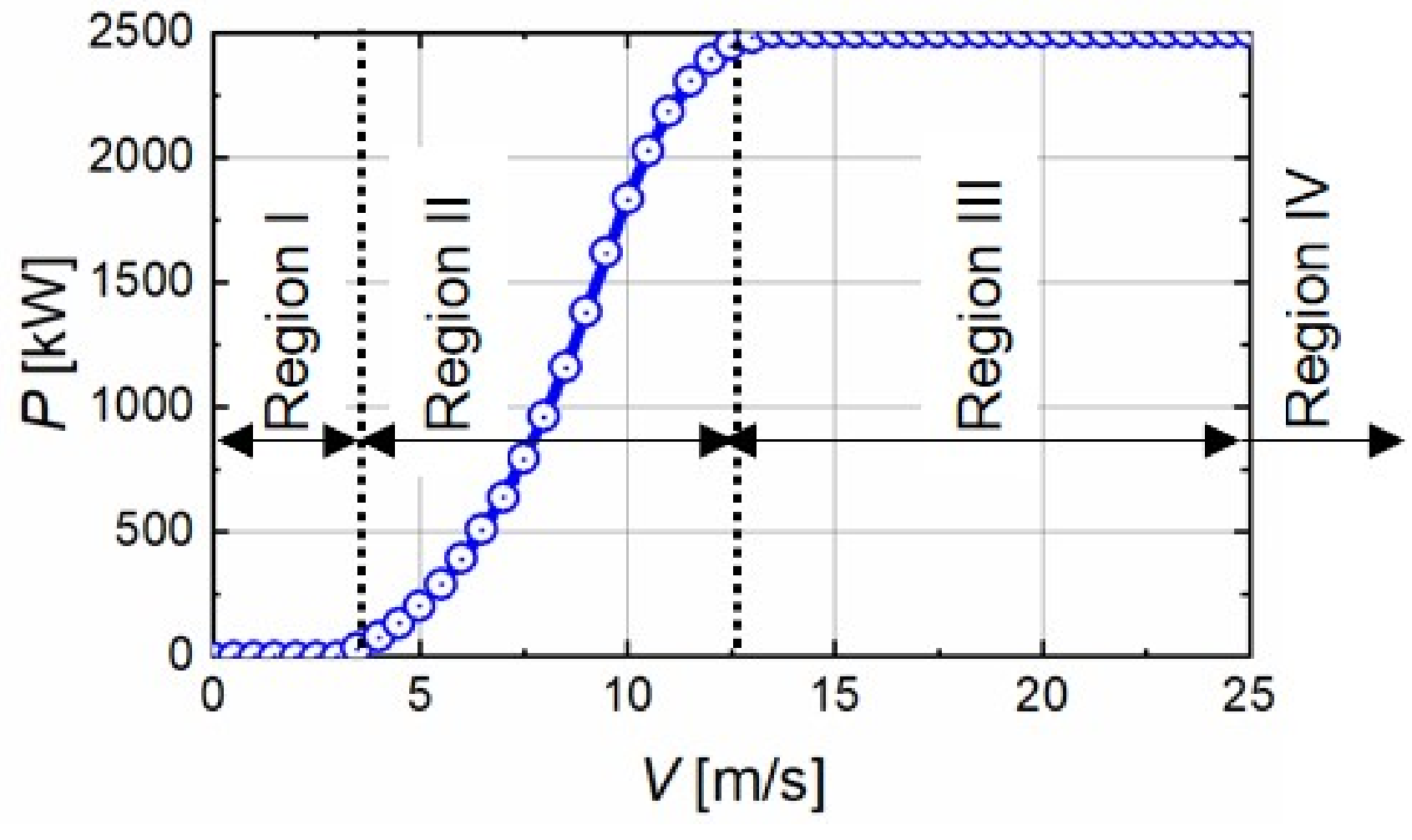

**Fig. 1**. Power curve of the 2.5 MW wind turbine at the Eolos station.

Icing loss refers to the difference between turbine measured power ($P_{\text{mea}}$) under icing conditions and corresponding power under no-icing conditions with the same inflow wind speed ($P_{\text{no-icing}}$). The latter one can be modeled via the wind speed ($V$) and turbine power curve shown in Fig. 1, using Eq. (1). Wind turbine yields zero power production in Region I ($V < V_{\text{cut-in}}$, cut-in wind speed) and Region IV ($V > V_{\text{cut-out}}$, cut-out wind speed). In Region III ($V_{\text{rated}} < V < V_{\text{cut-out}}$, rated wind speed), the turbine is locked at its maximum power. In Region II ($V_{\text{cut-in}} < V < V_{\text{rated}}$), the turbine power can be modeled using the wind speed data and polynomial fit of the power curve. Note that the ninth-order polynomial fit performs best among different-degree fittings in this region [23] and is used in the present study. It should be noted that we denote the modeled no-icing power as $P_{\text{no-icing}}$ and $\tilde{P}_{\text{no-icing}}$ for the cases using measured wind speed and forecasted wind speed from NWP, respectively.

$$P_{\text{no-icing}} = \begin{cases} 0 & \text{Region I} \\ c_9 V^9 + c_8 V^8 + \cdots + c_1 V + c_0 & \text{Region II} \\ P_{\text{max}} & \text{Region III} \\ 0 & \text{Region IV} \end{cases} \tag{1}$$

where $c_1, \ldots, c_9$ are the regression constants for the polynomial fit of the power curve, and $P_{\text{max}}$ represents the turbine maximum capacity.

### 2.2 Concepts of icing phases

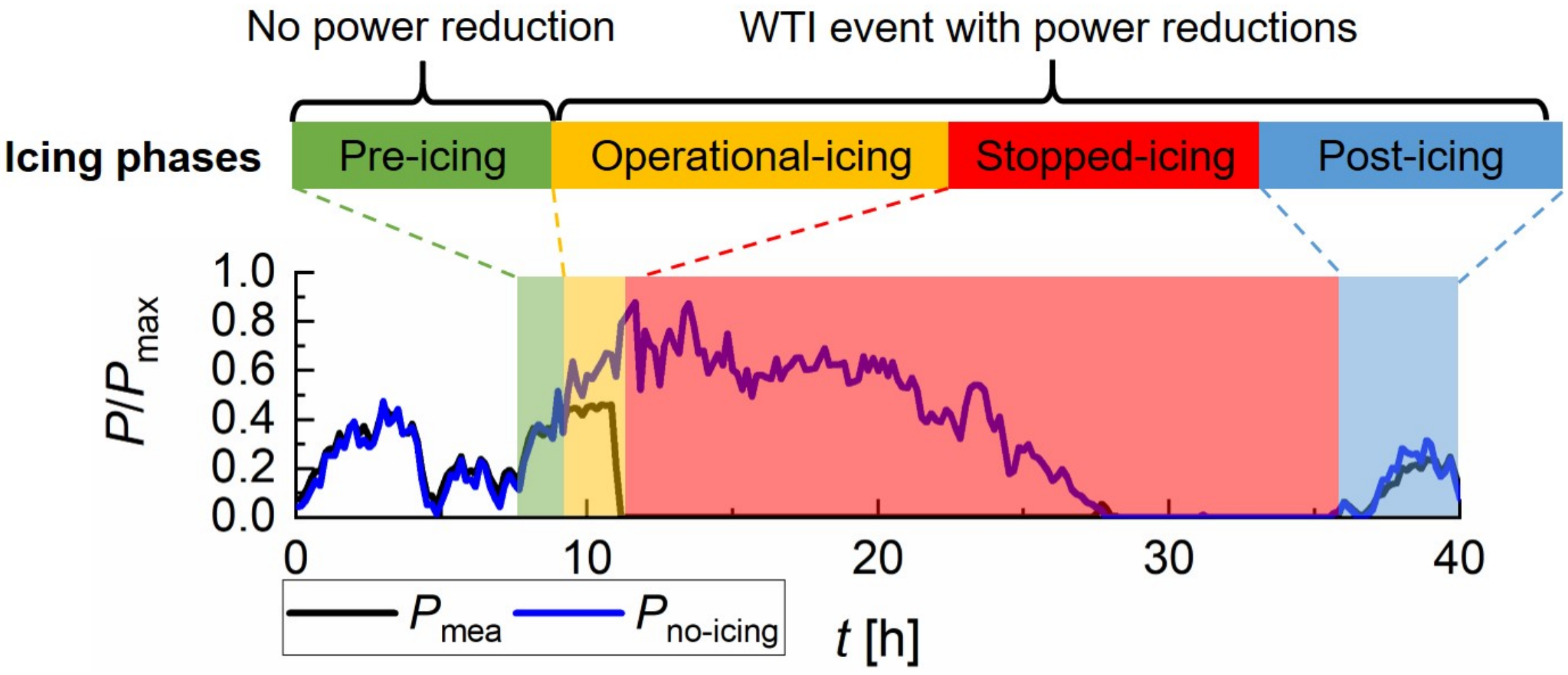

**Fig. 2**. Concept of icing phases of a severe WTI event (i.e., a sample retrieved from the Eolos database from 02/07/2019 00:05:00 to 02/18/2019 16:04:59)

The proposed statistical model is based on the various degrees of power losses in different icing phases. Our previous study [24] has demonstrated that a severe WTI event consists of four icing phases, i.e., pre-icing, operational-icing, stopped-icing, and post-icing phases. Pre-icing phase is an incubation period that wind turbine yields no power reduction subject to the meteorological icing phenomena, while the rest three phases exhibit detectable power losses (i.e., a threshold of 7.5%~20% [25], 15% used in the present study). As shown in Fig. 2, operational-icing phase refers to the period that a turbine starts yielding appreciable power losses until the turbine generates no power. In stopped-icing phase, a turbine is shut down or in the feathering status with no power generation due to the severe ice accretion. In the post-icing phase, a turbine resumes power production until it reaches full capacity associated with the slow natural ice melting process for the turbine without IPSs. In the present study, an entire duration of a WTI event is defined as the sum of the durations of the latter three icing phases with degraded power outputs. It should be emphasized that a turbine tends to yield the largest share of energy losses during the stopped-icing phase in a severe WTI event due to zero power generation and relatively long duration, as illustrated in Fig. 1. Based on such trend, we partition a WTI event into different icing phases for icing loss forecasts.

### 2.3 Cold climate effects

The cold climate (CC) effects on wind turbines can be grouped into three categories: precipitation icing (e.g., freezing rain/drizzle, snow, etc.), frost condensation (i.e., deposit in the form of ice crystal), and low-temperature effect (i.e., beyond the normal operating temperature range). Table 1 lists the detailed criteria for each effect. The most used meteorological icing criteria include air temperature $T < 0$°C and relative humidity $RH > 85\%$ [24,26,27] in the NWP variables, which are set as the base constraints for precipitation icing and frost condensation. Precipitation icing and frost condensation are negligible at extreme cold temperature below -30°C [28] and thus we use -30°C as the lower limit in these two categories. Most wind turbines can operate up to an extreme cold temperature of -20°C as suggested in the Recommendation Practice DNVGL-RP-0363 [29], while some turbines can maintain operation until -30°C. Here, we set a value of -30°C as the threshold for the data examination for low-temperature conditions for wind turbines. Besides the basic meteorological icing criteria, precipitation icing defined here also considers the periods with various kinds of precipitations that may cause turbine ice accretion, such as freezing rain/drizzle, snow, sleet, hail, mixture of rain and snow. To feed all the type of precipitations into our proposed model, binary information is used, i.e., 0: no precipitation and 1: with precipitations. As to the frost condensation, a threshold of 0.1 cm/h is set as the threshold to further filter out the frost condensation conditions that can contribute to detectable turbine ice accretion. It should be noted that precipitation icing, frost condensation, and low temperature have equivalent effects on turbines in CC regions. The condition that belongs to any of the three categories is to be counted into the duration estimation.

Table 1. Categorization criteria for three cold climate effects based on NWP variables.

| NWP variables | Precipitation icing | Frost condensation | Low temperature |
|---|---|---|---|
| $T$ [°C] | -30 ~ 0 | -30 ~ 0 | < -30 |
| $RH$ [%] | >85 | >85 | NA |
| Other feature | **OR** Precipitation index = 1 (Binary information 0: no precipitation; 1: with precipitations, i.e., rain and snow mixed, snow, sleet, freezing rain/drizzle, and hail) | **AND** Frost depth > 0.1 cm/h increase | NA |

### 2.4 Statistical wind turbine power loss model

The meteorological icing process determined by the weather conditions does not overlap perfectly with the wind turbine rotor icing process. There is usually an incubation period that is the time between the start of meteorological icing and the beginning of turbine rotor icing, dependent on the surface temperature of the turbine rotor [30]. The ice accretion process on the turbine surfaces occurs in the operational-icing and stopped-icing phases, followed by the post-icing phase. In the post-icing phase, the ice structures may remain persistent with no growth and no ablation. The ice structures may also be removed through ablation, includes melting, erosion, sublimation, and shedding of ice, which contributes to the delay between the end of meteorological icing and the end of turbine rotor icing. Here, the statistical icing model mainly bridges the gaps between meteorological and turbine icing in terms of the time/duration difference.

Fig. 3 illustrates the statistical icing model for the wind turbine power loss forecast. First, NWP variables listed in Table 1 (hourly resolution) are used as primary parameters to determine whether there is a potential WTI event that belongs to the above three categories. Specifically, the timestamps that satisfy the criteria of CC effects described in Table 1 are labeled. A threshold of 4 h is set to examine the timestamps, and the gaps between two labeled timestamps shorter than 4 h are eliminated. Such a 4-h threshold is also used by Davis [31] and he also pointed out this threshold has no appreciable difference with a 2-h threshold. If the consecutive timestamps make a window that lasts larger than 12 h, we assume that the CC effects can contribute to a severe WTI event. The start time and end time of the window (referred as to CC period hereafter) are denoted as $t_{\mathrm{CC,start}}$ and $t_{\mathrm{CC,end}}$. Note that in this section, the parameters with hat and without hat corresponding to the modeled and measured parameters, respectively.

Second, with a determined CC period, the start time ($\hat{t}_{\mathrm{WTI,start}}$) and end time ($\hat{t}_{WTI,end}$) of the potential WTI can be estimated using Eq. (2) and Eq. (3), respectively.

$$\hat{t}_{\mathrm{WTI,start}} = t_{\mathrm{CC,start}} + C_1 \quad (2)$$

where $C_1$ refers to the incubation factor that describes the delay of the onset of WTI event in comparison to the onset of the CC period, which can be calculated with $C_1 = a_1 \bar{V}_{\mathrm{CC}}$(unit: h, shift backward). $\bar{V}_{\mathrm{CC}}$ is the averaged wind speed during the CC period and $a_1 = 0.2$. The coefficient $a_1$ is derived by linear regression on the detected WTI events of a 2.5 MW turbine at Eolos Wind Energy Station at the University of Minnesota from 2011 to 2020 winters (i.e., October to next February). More detailed information of this database and the detection methods for WTI events using turbine SCADA signals are available in the literature [24].

$$\hat{t}_{\mathrm{WTI,end}} = C_2(t_{\mathrm{CC,end}} - t_{\mathrm{CC,start}}) + t_{\mathrm{CC,end}} \quad (3)$$

where $C_2$ is the persistence/ablation factors that corrects the delay between the end of CC period and the end of WTI event. It can be calculated using $C_2 = a_2 \Delta t_{\mathrm{abl}}$, where the ablation time ($\Delta t_{\mathrm{abl}}$ in a unit of h) is associated with the first time that the air temperature rises above 0°C ($t_{\mathrm{rec}}$), i.e., $\Delta t_{\mathrm{abl}} = t_{\mathrm{rec}} - t_{\mathrm{CC,end}}$. The coefficient $a_2$=0.02 is derived with linear regression from the Eolos database mentioned before. An upper band of 24 h is set in $\Delta t_{\mathrm{ablation}}$ to eliminate the overestimation in such delay. The total duration of a WTI event can then be calculated, as given in Eq. (4).

$$\widehat{D}_{\mathrm{WTI}} = \hat{t}_{\mathrm{WTI,end}} - \hat{t}_{\mathrm{WTI,start}} \quad (4)$$

Third, following the flowchart in Fig. 3, the WTI event is further partitioned into operational-icing, stopped-icing, and post-icing phases. The duration of the operational-icing phase ($\widehat{D}_{\mathrm{OP}}$) is proportional to the total duration of WTI with a correction factor of $C_3 = a_3/\bar{V}_{\mathrm{WTI}}^2$, where $\bar{V}_{\mathrm{WTI}}$ is the averaged wind speed during WTI event, as given in Eq. (5). The coefficient $a_3 = 8$ is derived with linear regression from the Eolos database like other coefficients mentioned earlier.

$$\widehat{D}_{OP} = C_3 \widehat{D}_{\mathrm{WTI}} \quad (5)$$

The duration of the stopped-icing phase ($\widehat{D}_{\mathrm{ST}}$) can be determined by the end time of the CC period, as shown in Eq. (6), while the duration of the post-icing phase ($\widehat{D}_{\mathrm{PO}}$) can be extracted using Eq. (7).

$$\widehat{D}_{\mathrm{ST}} = t_{\mathrm{CC,end}} - \hat{t}_{\mathrm{WTI,start}} - \widehat{D}_{\mathrm{OP}} \tag{6}$$

$$\widehat{D}_{\mathrm{PO}} = \widehat{D}_{\mathrm{WTI}} - \widehat{D}_{\mathrm{OP}} - \widehat{D}_{\mathrm{ST}} \tag{7}$$

Last, energy losses in three phases are modeled and added up for the total loss of the WTI event. Energy loss in the stopped-icing phase equals the energy supposed to be generated under the same inflows, as shown in Eq. (8).

$$\widehat{\Delta E}_{ST} = C_4 \sum \tilde{P}_{\mathrm{no-icing}} \Delta t \tag{8}$$

where $C_4$ is the scaling factor for different turbines, i.e., $C_4 = P_{\mathrm{max}}/P_{\mathrm{max,2.5MW}}$, if the power curve of a 2.5 MW turbine is used for the modeling of $\tilde{P}_{\mathrm{no-icing}}$ (Eq. (1) with forecasted wind speed from NWP). In the present study, $P_{\mathrm{max,2.5MW}}$ corresponds to the maximum capacity of a 2.5 MW turbine at the Eolos station, i.e., 2.5 MW. It should be noted that, in the cases that power curves for the target wind farms are available for $\tilde{P}_{\mathrm{no-icing}}$ calculation, $C_4$ equals 1 in Eq. (8). $\Delta t$ is the time interval of the datasets in a unit of h.

Energy losses in operational-icing phase ($\widehat{\Delta E}_{\mathrm{OP}}$) and post-icing phase ($\widehat{\Delta E}_{\mathrm{PO}}$) are proportional to their durations, as shown in Eq. (9) and Eq. (10).

$$\widehat{\Delta E}_{\mathrm{OP}} = C_4 C_5 \widehat{D}_{\mathrm{OP}} \tag{9}$$

$$\widehat{\Delta E}_{\mathrm{PO}} = C_4 C_6 \widehat{D}_{\mathrm{PO}} \tag{10}$$

where $C_4$ is the scaling factor for turbines with different capacity and it always equals to $P_{\mathrm{max}}/P_{\mathrm{max,2.5MW}}$ in Eqs. (9) and (10). $C_5$ and $C_6$ are correction factors associated with wind speeds, i.e., $C_5 = a_4 \bar{V}_{\mathrm{OP}}^2$, and $C_6 = a_5 \bar{V}_{\mathrm{WTI}}^2$, where $\bar{V}_{\mathrm{OP}}$ refers to the averaged wind speed in the operational-icing phase. The coefficients $a_4$ and $a_5$ have the same value of 15, which are derived with linear regressions from the Eolos database.

The total turbine energy loss during the WTI event can be calculated using Eq. (11).

$$\widehat{\Delta E}_{\mathrm{WTI}} = \widehat{\Delta E}_{\mathrm{OP}} + \widehat{\Delta E}_{\mathrm{ST}} + \widehat{\Delta E}_{\mathrm{PO}} \tag{11}$$

Such energy loss information can be readily integrated into the existing power forecasts ($\tilde{P}_{\mathrm{no-icing}}$) by introducing a sequence of reduction factors ($R$) during the WTI event period, as given in Eq. (12) for three icing periods.

$$\tilde{P}_{\mathrm{icing}} = R C_4 \tilde{P}_{\mathrm{no-icing}} \tag{12}$$

where $R_{\mathrm{ST}} = 0$ in the stopped-icing phase. In the operational-icing phase, $R_{\mathrm{OP}}$ is a descending sequence from 1 to 0 with a fixed step of $\Delta t/\widehat{D}_{\mathrm{OP}}$. In the post-icing phase, $R_{\mathrm{PO}}$ is an ascending sequence from 0 to 1 with a fixed step of $\Delta t/\widehat{D}_{\mathrm{PO}}$.

Moreover, the total energy loss in a wind farm ($\widehat{\Delta E}_{\mathrm{WTI,WF}}$) for the identified WTI event can be assessed with the turbine number information ($n$) using Eq. (13). If multiple WTI events are identified in the forecasting period (usually 1-10 days ahead), we need to sum the energy loss results for all WTI events to obtain the total energy loss in that period for the wind farm, as given in Eq. (14).

$$\widehat{\Delta E}_{\mathrm{WTI,WF}} = n \widehat{\Delta E}_{\mathrm{WTI}} \tag{13}$$

$$\widehat{\Delta E}_{\mathrm{WF}} = \sum \widehat{\Delta E}_{\mathrm{WTI(i),WF}} \quad (14)$$

where $i = 1, \ldots, m$ indicates the number of the WTI events in the forecasting period.

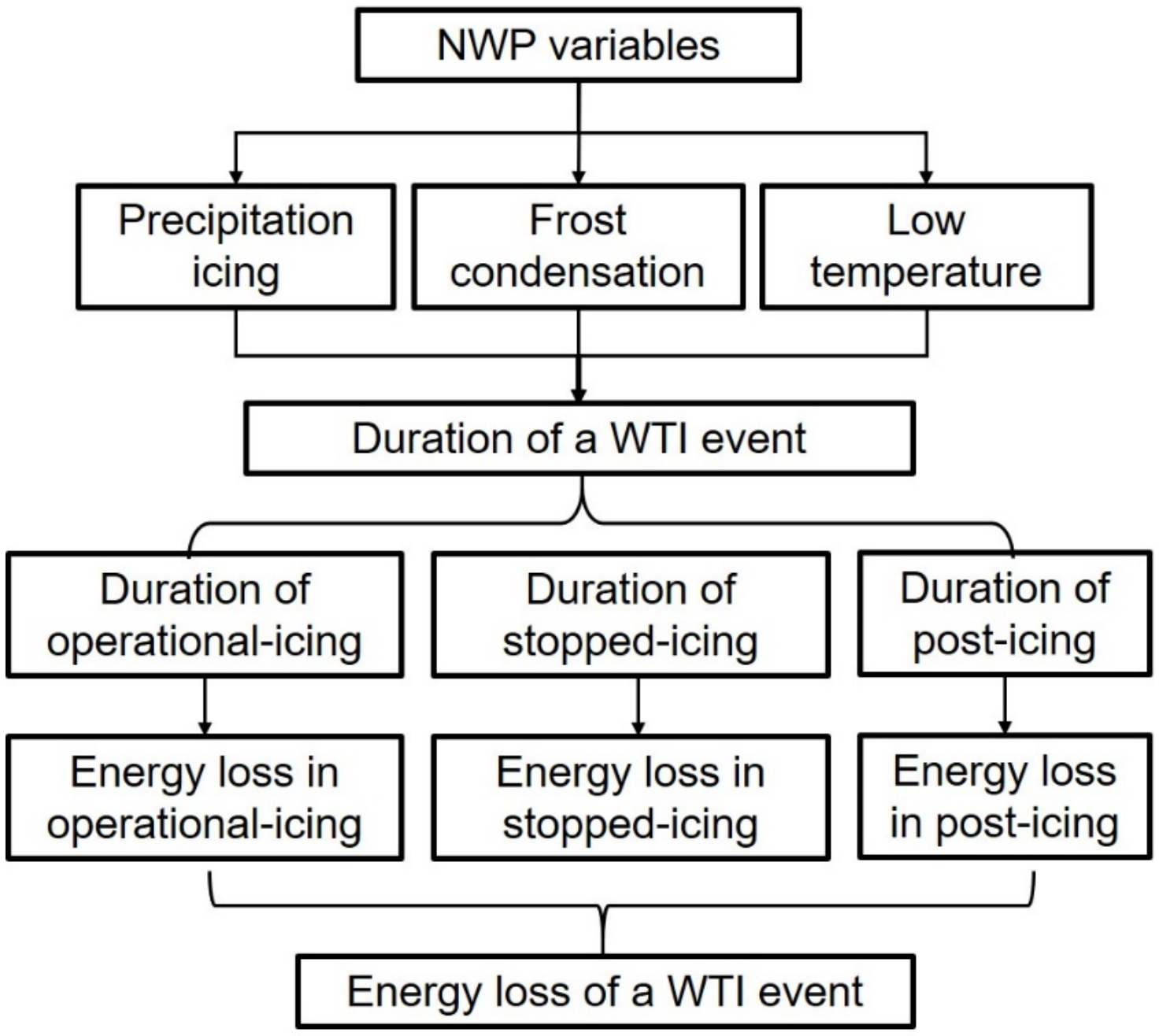

## 3. Model assessment at wind farm scale

The proposed statistical model is evaluated using the data from the three large-scale wind farms (i.e., capacity > 100 MW) across the United States, as listed in Table 2. The detailed information of turbine number and wind farm capacity are not revealed following the non-disclosure agreements with the wind farm operators. In general, the three wind farms are located in different regions in the United States and host turbines of different capacity and from different manufacturer, providing us a wide parameter space to assess the generalizability of our model derived from our Eolos database.

High-resolution NWP variables (temporal resolution: 1 h, spatial resolution: 90 m × 90 m) are used as inputs for the icing loss forecasts. The forecasts are further compared with the turbine SCADA data for performance evaluation. Table 3 summarizes the comparison between the measurements and forecasts in terms of icing duration and total energy loss for the identified icing events in the three wind farms. In general, the measured and forecasted durations of WTI events, the detailed onset time and end time of the WTI events are in good agreement for the three selected wind farms. In addition, the forecasted total wind farm energy losses are evaluated using a metric of relative mean absolute error, i.e., $\mathrm{RMAE} = (\widehat{\Delta E}_{\mathrm{WF}} - \Delta E_{\mathrm{WF}})/\Delta E_{\mathrm{WF}}$, where $\Delta E_{\mathrm{WF}}$ and $\widehat{\Delta E}_{\mathrm{WF}}$.are the measured and modeled total energy loss of the wind farm, respectively. The energy loss forecasts in Farm A, Farm B, Farm C yield $RMAE$s of 9%, 1% and 21%, respectively, indicating a good prediction performance. Considering the large variations in geographical locations and turbine capacities and manufacturers of these three wind farms, such results suggest a good generalizability of the proposed model.

Table 2. Key parameters of three large-scale wind farms for model assessment.

| Farm# | Turbine capacity [kW] | Turbine number | Farm capacity [MW] | Location | Test period |
|---|---|---|---|---|---|
| A | 1500 | >100 | >200 | Minnesota | 02/01/2019-02/05/2019 |
| B | 2000 | >50 | >100 | North Dakota | 01/28/2019-01/30/2019 |
| C | 2000 | >200 | >400 | Texas | 10/23/2020-10/28/2020 |

Table 3. Comparison of icing loss measurements and forecasts for the three selected wind farms.

| Parameter | Farm A | | Farm B | | Farm C | |
|---|---|---|---|---|---|---|
| | Measurement | Forecast | Measurement | Forecast | Measurement | Forecast |
| Duration [h] | 53 | 54 | 28 | 29 | 61 | 60 |
| Onset | 02/03/2019 22:10:00 | 02/03/2019 23:00:00 | 01/29/2019 06:50:00 | 01/29/2019 08:00:00 | 10/25/2020 23:30:00 | 10/26/2020 02:00:00 |
| End | 02/06/2019 03:10:00 | 02/06/2019 05:00:00 | 01/30/2019 12:00:00 | 01/30/2019 12:00:00 | 10/28/2020 12:50:00 | 10/28/2020 14:00:00 |
| Farm loss [GWh] | 3.7 | 3.4 | 1.9 | 1.9 | 18.8 | 14.9 |
| RMAE | 9% (-) | | 1% (+) | | 21% (-) | |

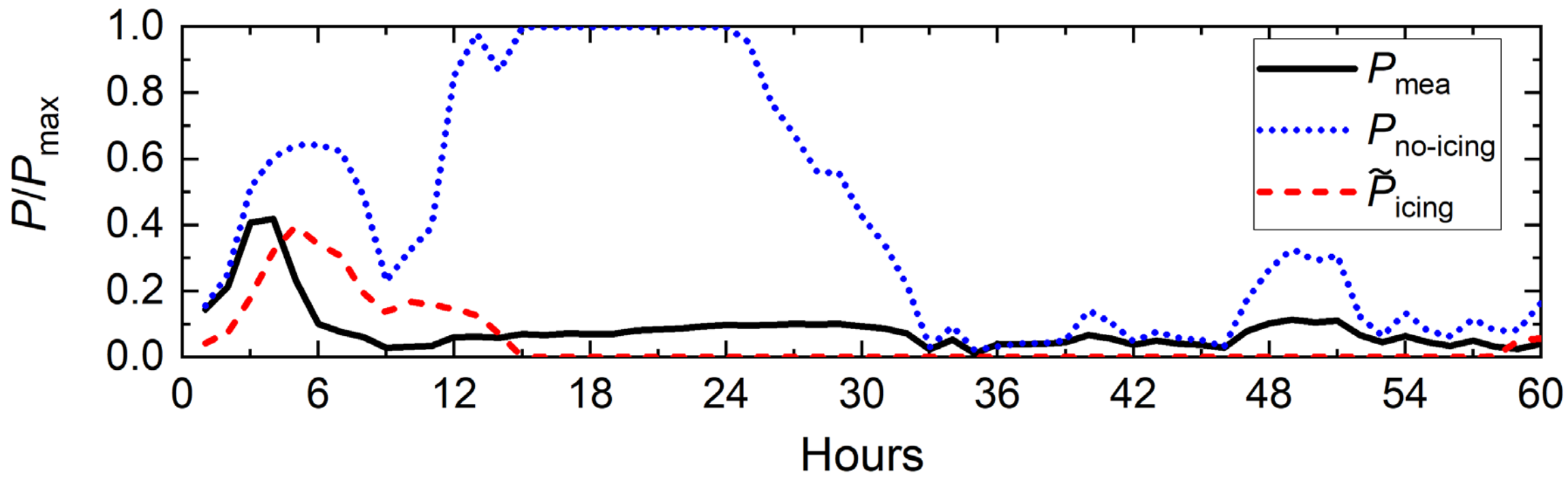

**Fig. 4**. Time series of the normalized turbine power by the turbine maximum capacity during the test period at Farm A starting from 02/03/2019 20:00:00, including measured power ($P_{\text{mea}}$), modeled power under the corresponding no-icing condition with measured wind speed ($P_{\text{no-icing}}$), and forecasted power with consideration of icing loss ($\tilde{P}_{\text{icing}}$). Note that the results here are the averaged data from all turbines in the wind farm.

Figs. 4-6 compare the time series of measured power $P_{\text{mea}}$, modeled power under a corresponding no-icing condition with measured wind speed $P_{\text{no-icing}}$, and forecast power with consideration of icing $\tilde{P}_{\text{icing}}$ for three wind farms. In general, the forecast of $\tilde{P}_{\text{icing}}$ can follow trend of the measurement of $P_{\text{mea}}$ in all three cases. Compared to the modeled power under the no-icing condition of $P_{\text{no-icing}}$, $\tilde{P}_{\text{icing}}$ yields much closer values with the real measurements in all three cases, suggesting the advantages of integrating icing losses into power forecasts. Nevertheless, deviations between $P_{\text{mea}}$ and $\tilde{P}_{\text{icing}}$ are more significant in the operational-icing phase compared to those in the stopped-icing and post-icing phases. Such trend is mainly due to the complex turbine

degraded operations in the operational-icing phase, which is challenging to be considered in wind farm or power system-level modeling. In addition, the case shown in Fig.5 shows an appreciable difference between $P_{\text{mea}}$ and $\tilde{P}_{\text{icing}}$. This is mainly due to the low accuracy of forecasted wind speed derived from the NWP system. Although this NWP system can generally predict wind speed with a deviation smaller than 1.0 m/s, it may still present some bad results in some applications [32]. Therefore, the uncertainties in the NWP data could also be contributors to such deviations.

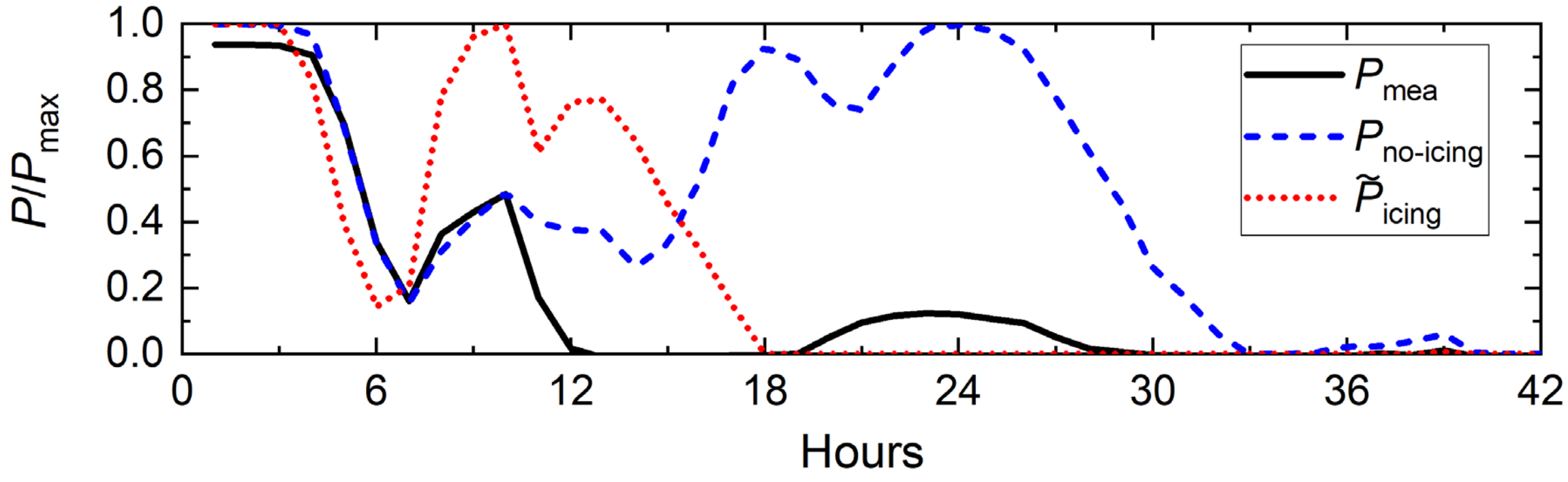

**Fig. 5**. Time series of normalized turbine power by the turbine maximum capacity during the test period at Farm B starting from 01/28/2019 21:00:00, including measured power ($P_{\text{mea}}$), modeled power under the corresponding no-icing condition with measured wind speed ($P_{\text{no-icing}}$), and forecasted power with consideration of icing loss ($\tilde{P}_{\text{icing}}$). Note that the results here are the averaged data from all turbines in the wind farm.

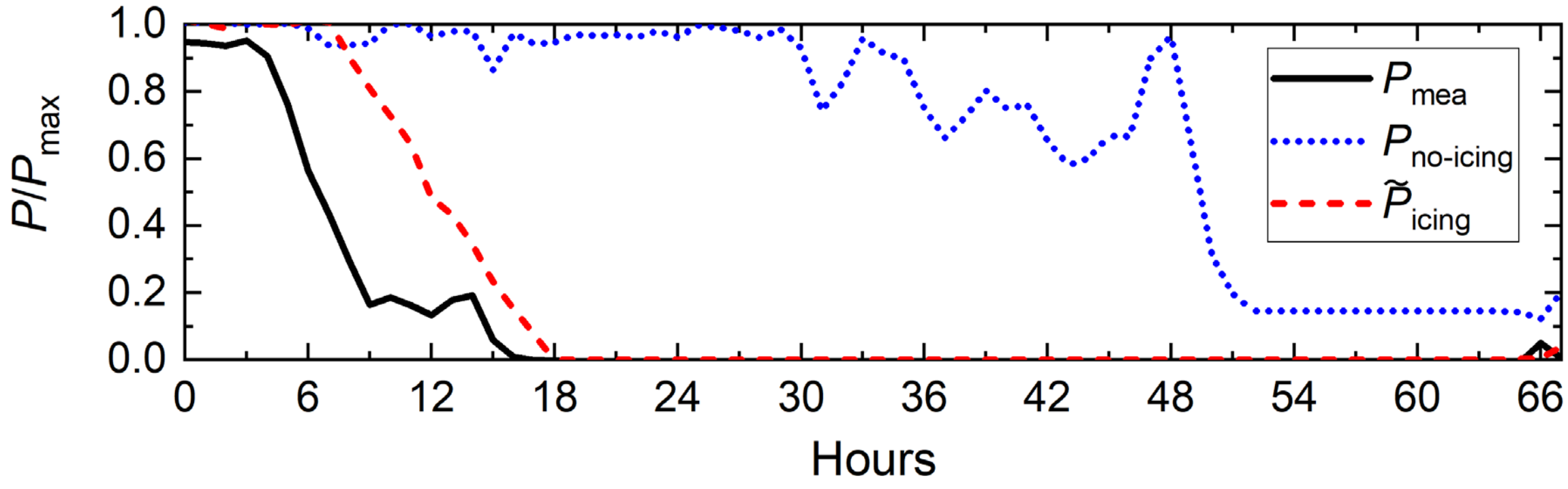

**Fig. 6**. Time series of normalized turbine power by the turbine maximum capacity during the test period at Farm C starting from 10/25/2020 19:00:00, including measured power ($P_{\text{mea}}$), modeled power under the corresponding no-icing condition with measured wind speed ($P_{\text{no-icing}}$), and forecasted power with consideration of icing loss ($\tilde{P}_{\text{icing}}$). Note that the results here are the averaged data from all turbines in the wind farm.

## 4. Model assessment at the power system scale

The proposed statistical icing loss model only depends on the NWP variables and basic wind farm information, such as turbine number, turbine capacity, one turbine power curve, and has no requirements of the turbine SCADA data. The model can be implemented for power system-level forecasts over a large geographic region. This section will implement the proposed model to 143 large-scale wind farms (>100 MW) in Texas for an hourly one-week ahead forecast from 02/08/2021 12:00:00 to 02/15/2021 12:00:00 local time. A high-resolution NWP database (temporal resolution: 1 h, spatial resolution: 90 m × 90 m) is used, the same as the one used in Section 3. The farm capacity ranges from 100 MW to 525 MW, while the turbine capacity ranges from 1.0 to 3.6 MW. The total capacity of these 143 wind farms is 28.29 GW, corresponding to 91% of the total wind capacity in Texas. The total number of turbines is 13852, representing 83.3% of the total installation number. Detailed information of these wind farms in terms of location, turbine capacity, turbine number, etc., are listed in Table A1 in Appendix.

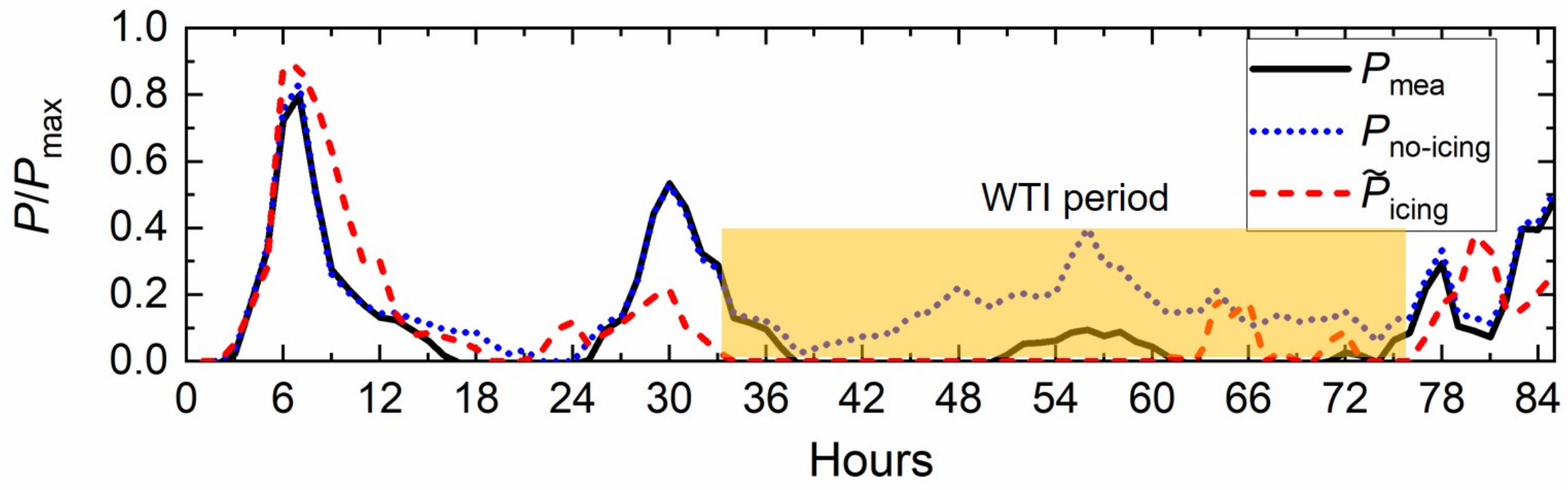

**Fig. 7**. Time series of normalized turbine power by the turbine maximum capacity in the period from 02/08/2021 12:00:00 to 02/11/2021 23:00:00 local time at Farm C in Texas, including measured power ($P_{\mathrm{mea}}$), modeled power under the corresponding no-icing condition with measured wind speed ($P_{\mathrm{no-icing}}$), and forecasted power with consideration of icing loss ($\tilde{P}_{\mathrm{icing}}$). Yellow square highlights the detected WTI period via the farm measurements. Note that the results here are the averaged data from all turbines in the wind farm.

Fig. 7 shows the prediction results during the forecasting period for one of the 143 wind turbines, i.e., Farm C in Table 2, and compares the prediction with farm measurement data. A 36-h WTI event is observed from 02/10/2021 01:00:00 to 02/11/2021 13:00:00 local time, as highlighted by yellow in the figure. The statistical model successfully predicts the occurrence of this WTI event and forecasts a 42-h WTI event from 02/09/2021 15:00:00 to 02/11/2021 09:00:00 local time. The total farm energy loss for the identified WTI event is slightly overestimated by 15.7% with the statistical model, i.e., 2.95 GWh (forecasted) vs. 2.55 GWh (measured). This case further validates the capability of the proposed model for further analysis.

Fig. 8 shows the prediction results for all 143 wind farms in Texas regarding absolute energy loss values and corresponding risk levels. The energy loss per farm ranges from 0.24 GW to 28.32 GW across 143 wind farms for one week. We further estimate the farm risk level with the ratio of forecasted total energy loss over the forecasted total energy production under no-icing conditions. There are 42 wind farms rated in low risk (L1: 0~20% energy loss), while 41 wind farms are at the level of extremely high risk (L5: 80~100% energy loss). The numbers of wind farms in moderate risk (L2: 20~40% energy loss), moderate-high risk (L4: 40~60% energy loss), and high risk (L5: 60~80% energy loss) are 23, 22, 15, respectively. In Fig. 8, most wind farms rating with high-risk levels (i.e., L3-L5) are concentrated in the northwest of Texas, while the wind farms located in the southern part of Texas have relatively low risks in terms of energy loss. This trend matches well with the Arctic airmass spreading direction across Texas from 02/11/2021 to 02/20/2021 [33]. Specifically, a cluster of wind farms along the boundary of two Köppen climate classifications (i.e., left: cold semi-arid, right: humid subtropical)[34] are found to be rated in L5: extreme high risk, suggesting the wind farm icing losses are highly associated with the local weather and geographic conditions. Local-dependent weather and topographic features could be rendering certain wind farms a lower risk in the northwest while the majority are at relatively high-risk levels. Such discrepancies are also observed from one of the wind farms in the south, which suffers from L5: extreme high risk while the majority are at the relatively low risk level, which might be associated with the fact that this wind farm is located closer to the hot semi-arid and humid subtropical boundary. It should be cautioned that complex terrain or human activities (e.g., urban heat island) could also decrease the accuracy of NWP data and thus affect the ranking results.

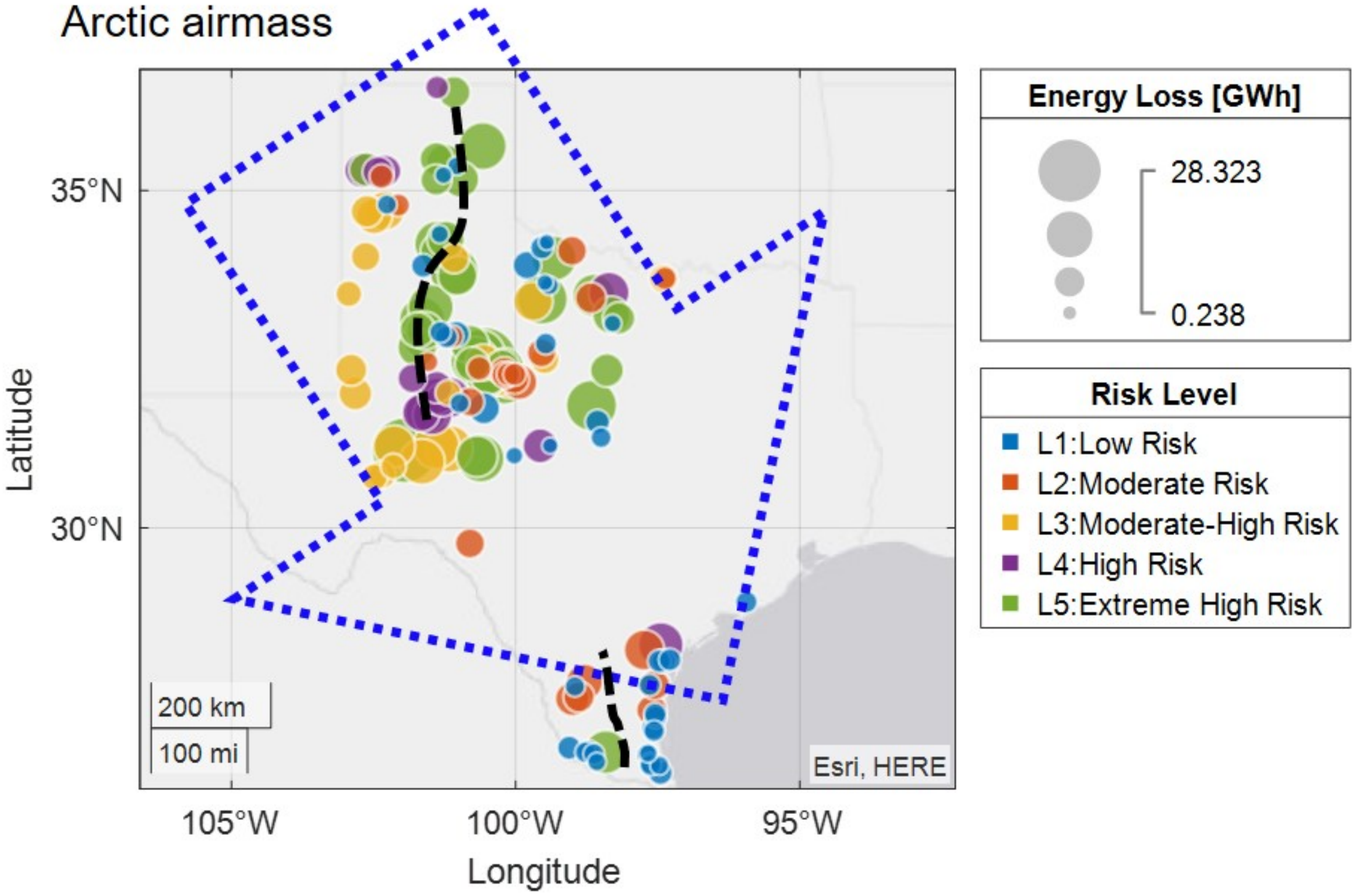

**Fig. 8**. Forecasts of energy loss in 143 large-scale wind farms in Texas and corresponding risk levels, including L1: low risk (0~20% energy loss), L2: moderate risk (20~40% energy loss), L3: moderate-high risk (40~60% energy loss), L4: high risk (60~80% energy loss) and L5: extreme high risk (80~100% energy loss). The size of the circle corresponds to the absolute value of the energy loss for each wind farm. The arrow indicates the deadly Arctic airmass spreading direction across Texas from 02/11/2021 to 02/20/2021. The black dash lines sketch out boundaries of regions belonging to different Köppen climate classifications.

Fig.9(a) compares the power forecasts without consideration of icing-induced power losses ($\tilde{P}_{\text{no-icing}}$) and the power forecasts with consideration of the icing-induced power losses via our proposed statistical model ($\tilde{P}_{\text{icing}}$) for 143 wind farms in Texas during the period from 02/08/2021 12:00:00 to 02/15/2021 12:00:00 local time. Fig. 9(b) provides the corresponding ratios of such two types of power forecasts (i.e., $\tilde{P}_{\text{icing}}/\tilde{P}_{\text{no-icing}}$). An appreciable deviation between the absolute values of $\tilde{P}_{\text{no-icing}}$ and $\tilde{P}_{\text{icing}}$ is observed since the morning of 02/09/2021 in Fig. 9(a). The deviation peaks on the morning of 02/14/2021 with a value of 13176 MW based on our prediction. Such value is very close to the reported forced outages of 14000 MW of wind (with a share of 24.8% of total installation capacity) and solar (with a share of 3.8% of the total installation capacity) by the Electric Reliability Council of Texas (ERCOT) [33], suggesting the proposed model is capable of accurately predicting the offline power for the power system. In Fig.9(b), three periods with over 50% outages (i.e., $\tilde{P}_{\mathrm{icing}}/\tilde{P}_{\mathrm{no-icing}} < 0.5$) are highlighted in red, i.e., 17 h from 02/10/2021 to 02/11/2021 (period 1), 3 h at midnight on 02/12/2021 (period 2), and 37 h from 02/13/2021 to 02/15/2021 (period 3). Such a 50% outage level approaches a threat to the power system with a high penetration of wind energy [33]. Period 1 corresponds to the beginning time of the Arctic cold covering nearly half of the state [33]. Period 3 lasts 37 h, and with the increasing load demand from customers on 02/15/2021 (Monday), ERCOT initiated rolling outage across much of the state of Texas [33].

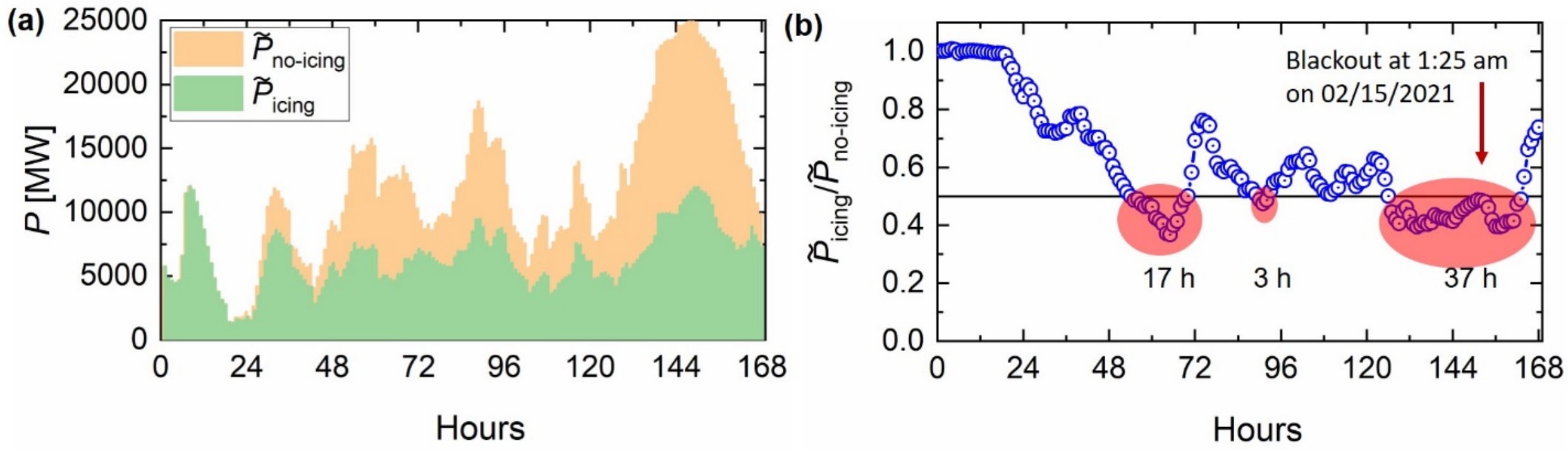

**Fig. 9**. Time evolution of the forecasted overall power output from 143 wind farms under no-icing conditions and icing conditions (a) and the corresponding ratio between them (b), starting from 02/08/2021 12:00:00. Red circles in Fig. 9(b) highlight the periods where power outputs are below half of the supposed productions.

The case for the icing loss forecasts in the course of the rolling outage periods in Texas in February 2021 suggests that the proposed forecast model is capable of providing accurate icing loss prediction at the power system level. Considering the reduced accuracy of NWP data as a function of forecast time, we recommend implementing the proposed model for the hourly-based prediction of energy loss only one-week ahead of a potential WTI event. In the present study, the baseline forecast of power under no-icing conditions is based on the wind speed data from NWP and the turbine power curve. This deterministic baseline power forecast approach can be replaced by state-of-the-art probabilistic models [16,35,36]. Our proposed statistical model can be readily integrated into the final forecasts as add-on corrections for the CC effects. With the more accurate energy offline forecasts, power system operators can develop more appropriate and pinpointed plans to balance the severe and sudden energy deficits and increase the system integrity under winter extremes.

## 4. Conclusions

In the present study, we propose a fast and robust statistical model to forecast wind farm energy losses under winter extremes. Specifically, this model uses high-resolution numerical weather prediction (NWP) variables to estimate the duration of a wind turbine icing (WTI) event caused by different types of icing phenomena, including precipitation icing, frost condensation, and low-temperature impact. This model further partitions a WTI event into operational-icing, stopped-icing, and post-icing phases for power loss estimation and then obtains the total energy loss during the entire WTI event. The main advantages of the statistical model are summarized below.

- The model can achieve a fast one-week-ahead forecast of the occurrence of a severe WTI event, the onset time and end time of the WTI event, and icing-induced energy loss.
- The model can be readily integrated into the existing power forecast models/systems that lack consideration of cold climate (CC) effects (i.e., icing and low temperature).
- The model has minimal requirements for the data input, i.e., only the most commonly-used NWP variables and basic information of a wind farm (e.g., turbine number and capacity) are needed without the need for turbine SCADA data.
- The model can be implemented to the power system-level forecasts, which can assist system operators in establishing appropriate and timely emergent plans for winter extremes.

The proposed model has been evaluated at both wind farm and power system scales. At the wind farm scale, the model can successfully predict the WTI events and the corresponding energy losses with errors less than 25% based on the assessments conducted at three large-scale wind farms (> 100 MW) across the United States. At the power system level, the model is used to forecast the energy losses from 143 wind farms (~91% of total wind capacity) in Texas during the rolling outage periods in February 2021. The prediction of the total energy loss from these wind farms is 13176 MW in the morning of 02/14/2021, which is in good agreement with the forced outages of 14000 MW of wind (with a share of 24.8% of total installation capacity) and solar (with a share of 3.8% of the total installation capacity) by Electric Reliability Council of Texas (ERCOT) [33].

In the end, we would like to caution the readers that our statistical model is designed for utility-scale wind turbines (>1 MW), and its validity for small-scale turbine icing loss forecast has not been assessed and is outside the scope of our current study. In addition, our model highly depends on the accuracy of NWP variables. The influence of different input variables with respect to their accuracy on prediction performance needs to be comprehensively investigated in the future. More validation work also needs to be conducted to wind farms located in complex terrains, offshore sites, and outside the United States to improve the robustness and reliability of the proposed model. Moreover, uncertainty analysis will be conducted to further optimize the current model for probabilistic forecasts. Furthermore, modified versions of the proposed model pertinent to the wind turbines equipped with various anti/de-icing systems [37,38] or vertical axis turbines [39] are also demanded.

## Appendix

Table A1. Key parameters of 143 wind farms in Texas for the icing loss forecast in February 2021.

| Farm name | Farm capacity [MW] | Turbine number | Turbine capacity [kW] | Hub height [m] | Rotor diameter [m] | Longitude | Latitude |
|---|---|---|---|---|---|---|---|
| 'Amazon Wind Farm Texas' | 253 | 110 | 2300 | 80 | 116 | -101.04 | 32.91 |
| 'Aviator Wind' | 525.02 | 191 | 2720 | 90 | 116 | -100.56 | 31.81 |
| 'Baffin' | 202 | 101 | 2000 | 90 | 97 | -97.61 | 27.19 |
| 'Barton Chapel' | 120 | 60 | 2000 | 78 | 87 | -98.29 | 33.07 |
| 'Bearkat I' | 196.65 | 57 | 3450 | 87 | 126 | -101.58 | 31.75 |
| 'Bearkat II' | 162.15 | 47 | 3450 | 87 | 126 | -101.66 | 31.74 |
| 'Bethel' | 276 | 120 | 2300 | 80 | 116 | -102.49 | 34.63 |
| 'Blue Cloud I' | 148.35 | 43 | 3450 | 87 | 126 | -102.64 | 34.05 |
| 'Blue Summit' | 135.4 | 85 | 1600 | 80 | 82.5 | -99.45 | 34.25 |
| 'Blue Summit III' | 200.2 | 82 | 2520 | 89 | 127 | -99.54 | 34.17 |
| 'Bobcat Bluff' | 150 | 100 | 1620 | 80 | 91 | -98.63 | 33.48 |
| 'Brazos Wind Ranch' | 160 | 160 | 1000 | 60 | 61.4 | -101.33 | 32.94 |
| 'Briscoe' | 150 | 81 | 1850 | 80 | 87 | -101.34 | 34.37 |
| 'Bruenning's Breeze' | 228 | 76 | 3000 | 87.5 | 125 | -97.67 | 26.51 |
| 'Buckthorn' | 100.05 | 29 | 3450 | 87 | 126 | -98.40 | 32.37 |
| 'Buffalo Gap 2' | 232.5 | 155 | 1500 | 80 | 77 | -100.20 | 32.36 |
| 'Buffalo Gap 3' | 170.2 | 74 | 2300 | 80 | 93 | -100.19 | 32.30 |
| 'Buffalo Gap Wind Farm' | 120.6 | 67 | 1800 | 78 | 80 | -100.13 | 32.31 |
| 'Bull Creek' | 180 | 180 | 1000 | 69 | 61.4 | -101.60 | 32.96 |
| 'Cactus Flats' | 148.35 | 43 | 3450 | 87 | 126 | -100.02 | 31.10 |
| 'Callahan Divide' | 114 | 76 | 1500 | 80 | 87 | -100.02 | 32.31 |
| 'Cameron' | 165 | 55 | 3000 | 87.5 | 125 | -97.45 | 26.20 |
| 'Camp Springs' | 130.5 | 87 | 1500 | 80 | 77 | -100.81 | 32.76 |
| 'Canadian Breaks' | 200.1 | 87 | 2300 | 80 | 108 | -102.36 | 35.20 |
| 'Capricorn Ridge' | 364 | 208 | 1500 | 80 | 91 | -100.81 | 31.90 |
| 'Capricorn Ridge expansion' | 298.5 | 199 | 1500 | 80 | 87 | -100.99 | 31.88 |
| 'Cedro Hill' | 150 | 100 | 1500 | 80 | 82.5 | -98.96 | 27.55 |
| 'Champion' | 126.5 | 55 | 2300 | 80 | 93 | -100.64 | 32.40 |
| 'Chapman Ranch' | 249.08 | 81 | 3075 | 87.5 | 125 | -97.55 | 27.58 |
| 'Cranell' | 220 | 100 | 2200 | 92 | 120 | -97.45 | 28.20 |
| 'Desert Sky' | 129 | 86 | 1500 | 65 | 82.5 | -102.15 | 30.93 |
| 'Elbow Creek Wind Farm' | 121.9 | 53 | 2300 | 80 | 108 | -101.40 | 32.12 |
| 'Electra Wind' | 230 | 100 | 2300 | 80 | 116 | -99.01 | 34.13 |
| 'Falvez Astra' | 163.2 | 68 | 2400 | 80 | 107 | -102.06 | 34.79 |
| 'Flat Top' | 200 | 100 | 2000 | 95 | 110 | -98.56 | 31.60 |
| 'Fluvanna I' | 155.4 | 74 | 2100 | 80 | 114 | -101.12 | 32.87 |
| 'Foard City' | 350.28 | 139 | 2520 | 89 | 127 | -99.80 | 33.92 |
| 'Forest Creek Wind Project' | 124.2 | 54 | 2300 | 80 | 93 | -101.19 | 32.04 |
| 'Goldthwaite' | 141.1 | 83 | 1700 | 80 | 100 | -98.50 | 31.37 |
| 'Gopher Creek Wind Farm' | 158 | 79 | 2000 | 94 | 116 | -101.22 | 32.87 |
| 'Grandview I' | 211.22 | 118 | 1790 | 80 | 100 | -101.27 | 35.22 |
| 'Grandview II (Colbeck's Corner)' | 200.48 | 112 | 1790 | 80 | 100 | -101.04 | 35.36 |
| 'Green Pastures I' | 150 | 50 | 3000 | 92 | 116 | -99.41 | 33.62 |
| 'Green Pastures II' | 150 | 50 | 3000 | 92 | 116 | -99.48 | 33.67 |
| 'Gulf Wind' | 283.2 | 118 | 2400 | 80 | 92 | -97.58 | 26.91 |
| 'Gunsight' | 119.93 | 67 | 1790 | 80 | 100 | -101.54 | 32.50 |
| 'Hackberry' | 165.6 | 72 | 2300 | 80 | 93 | -99.47 | 32.77 |
| 'Hale Wind' | 478 | 239 | 2000 | 94 | 116 | -101.63 | 33.91 |
| 'Heart of Texas' | 179.88 | 64 | 2820 | 89 | 127 | -99.39 | 31.25 |
| 'Hereford 1' | 199.9 | 104 | 2000 | 95 | 100 | -102.26 | 34.80 |
| 'Hidalgo' | 150 | 75 | 2000 | 80 | 110 | -98.41 | 26.53 |
| 'High Lonesome' | 449.94 | 142 | 3465 | 85 | 132 | -101.96 | 31.17 |
| 'Horse Creek Wind' | 230 | 100 | 2300 | 80 | 116 | -99.53 | 33.43 |
| 'Horse Hollow Expansion' | 210 | 140 | 1500 | 80 | 87 | -100.03 | 32.24 |
| 'Horse Hollow II' | 299 | 130 | 2300 | 80 | 108 | -99.93 | 32.20 |
| 'Horse Hollow III' | 223.5 | 149 | 1500 | 80 | 87 | -100.31 | 32.30 |
| 'Horse Hollow Wind Energy Center' | 210 | 140 | 1500 | 80 | 87 | -100.11 | 32.26 |
| 'Inadale (Roscoe IV)' | 197 | 197 | 1000 | 69 | 61.4 | -100.57 | 32.49 |
| 'Javelina' | 249.69 | 126 | 2000 | 94 | 116 | -99.00 | 27.36 |
| 'Javelina II' | 200 | 100 | 2000 | 94 | 116 | -98.89 | 27.39 |
| 'Jumbo Hill' | 160.74 | 57 | 2820 | 89 | 127 | -102.89 | 32.38 |
| 'Jumbo Road' | 299.7 | 162 | 1850 | 80 | 87 | -102.30 | 34.71 |
| 'Karankawa' | 307.46 | 124 | 2520 | 89 | 127 | -97.74 | 28.12 |
| 'Keechi' | 110 | 55 | 2000 | 95 | 100 | -98.18 | 33.14 |
| 'King Mtn. Wind Ranch' | 278.2 | 214 | 1300 | 60 | 65 | -102.14 | 31.24 |
| 'Langford' | 150 | 100 | 1500 | 80 | 77 | -100.68 | 31.09 |
| 'Live Oak' | 199.5 | 76 | 2625 | 85 | 120 | -100.62 | 31.05 |
| 'Lockett' | 183.75 | 75 | 2450 | 89 | 127 | -99.33 | 34.02 |
| 'Logan's Gap Wind' | 200.1 | 87 | 2300 | 80 | 108 | -98.67 | 31.85 |
| 'Lone Star I' | 200 | 100 | 2000 | 78 | 83 | -99.55 | 32.63 |
| 'Lone Star II' | 156 | 78 | 2000 | 78 | 87 | -99.49 | 32.53 |
| 'Longhorn' | 200 | 100 | 2000 | 80 | 100 | -101.23 | 34.29 |
| 'Loraine' | 100.5 | 67 | 1500 | 80 | 77 | -100.77 | 32.46 |
| 'Los Vientos I' | 200.1 | 87 | 2300 | 98.1 | 108 | -97.63 | 26.32 |
| 'Los Vientos II' | 201.6 | 84 | 2400 | 90 | 102 | -97.64 | 26.35 |
| 'Los Vientos III' | 200 | 100 | 2000 | 95 | 110 | -98.63 | 26.50 |
| 'Los Vientos IV' | 200 | 100 | 2000 | 95 | 110 | -98.71 | 26.53 |
| 'Los Vientos V' | 110 | 55 | 2000 | 95 | 110 | -98.57 | 26.37 |
| 'Magic Valley' | 203.28 | 112 | 1815 | 80 | 100 | -97.68 | 26.49 |
| 'Mariah North Wind' | 230.4 | 96 | 2400 | 80 | 107 | -102.62 | 34.70 |
| 'McAdoo' | 150 | 100 | 1500 | 80 | 77 | -101.03 | 33.77 |
| 'Mesquite Creek' | 211.22 | 118 | 1790 | 80 | 100 | -101.74 | 32.71 |
| 'Mesquite Star' | 418.9 | 118 | 3550 | 101.5 | 132 | -100.55 | 32.53 |
| 'Meste' | 201.6 | 56 | 3600 | 112 | 136 | -98.77 | 26.52 |
| 'Miami' | 288.6 | 156 | 1850 | 80 | 87 | -100.58 | 35.64 |
| 'Midway' | 162.86 | 47 | 3465 | 84 | 132 | -97.29 | 27.97 |
| 'Noble Great Plains' | 114 | 76 | 1500 | 80 | 77 | -101.38 | 36.47 |
| 'Notrees' | 152.61 | 96 | 1650 | 80 | 82 | -102.82 | 32.03 |
| 'Old Settler Wind' | 151.2 | 63 | 2400 | 80 | 107 | -101.08 | 34.02 |
| 'Palmas Altas' | 144.9 | 46 | 3150 | 87.5 | 125 | -97.47 | 26.32 |
| 'Palo Duro' | 249.9 | 147 | 1700 | 80 | 100 | -101.08 | 36.41 |
| 'Panhandle Wind 1' | 218.3 | 118 | 1850 | 80 | 87 | -101.25 | 35.41 |
| 'Panhandle Wind 2' | 181.7 | 79 | 2300 | 80 | 108 | -101.40 | 35.46 |
| 'Panther Creek I' | 142.5 | 95 | 1620 | 80 | 87 | -101.40 | 31.99 |
| 'Panther Creek II' | 115.5 | 77 | 1620 | 80 | 87 | -101.30 | 31.93 |
| 'Panther Creek III' | 199.5 | 133 | 1500 | 80 | 77 | -101.16 | 31.97 |
| 'Papalote Creek' | 179.85 | 109 | 1650 | 80 | 82 | -97.47 | 27.95 |
| 'Papalote Creek II' | 200.1 | 87 | 2300 | 80 | 101 | -97.40 | 27.95 |
| 'Patriot Wind' | 226.05 | 63 | 3600 | 82 | 136 | -97.64 | 27.58 |
| 'Penescal' | 201.6 | 84 | 2400 | 80 | 92 | -97.54 | 27.07 |
| 'Penescal II' | 201.6 | 84 | 2400 | 80 | 92 | -97.55 | 27.12 |
| 'Peyton Creek' | 151.2 | 48 | 3150 | 87.5 | 125 | -95.94 | 28.87 |
| 'Pyron (Roscoe III)' | 249 | 166 | 1500 | 80 | 77 | -100.64 | 32.59 |
| 'Ranchero' | 300 | 120 | 2500 | 89 | 127 | -101.65 | 31.00 |
| 'Rattlesnake' | 207 | 118 | 1790 | 80 | 100 | -101.47 | 31.70 |
| 'Rattlesnake Wind' | 160 | 64 | 2500 | 90 | 109 | -99.57 | 31.24 |
| 'Rio Bravo' | 237.6 | 66 | 3600 | 105 | 136 | -99.06 | 26.60 |
| 'Rocksprings' | 149.34 | 69 | 2300 | 80 | 116 | -100.80 | 29.76 |
| 'Roscoe' | 209 | 209 | 1000 | 69 | 61.4 | -100.72 | 32.51 |
| 'Route 66' | 150 | 75 | 2000 | 80 | 110 | -101.40 | 35.16 |
| 'Sage Draw' | 338.4 | 120 | 2820 | 89 | 127 | -101.50 | 33.30 |
| 'Salt Fork' | 174 | 87 | 2000 | 80 | 100 | -100.97 | 35.16 |
| 'Santa Rita' | 300 | 120 | 2500 | 90 | 116 | -101.40 | 31.25 |
| 'Santa Rita East' | 302.4 | 120 | 2520 | 89 | 127 | -101.17 | 31.20 |
| 'Senate' | 150 | 75 | 2000 | 100 | 90 | -98.32 | 33.19 |
| 'Shannon' | 204.09 | 119 | 1715 | 80 | 103 | -98.36 | 33.52 |
| 'Sherbino I' | 150 | 50 | 3000 | 80 | 90 | -102.35 | 30.81 |
| 'Sherbino II' | 150 | 60 | 2500 | 80 | 96 | -102.49 | 30.77 |
| 'South Plains' | 200 | 100 | 2000 | 80 | 100 | -101.33 | 34.11 |
| 'South Plains II' | 300.3 | 91 | 3300 | 91.5 | 117 | -101.37 | 34.23 |
| 'South Trent Mesa' | 101.2 | 44 | 2300 | 80 | 82 | -100.24 | 32.42 |
| 'Spinning Spur 2' | 160.95 | 87 | 1850 | 80 | 87 | -102.63 | 35.31 |
| 'Spinning Spur 3' | 194 | 97 | 2000 | 80 | 100 | -102.71 | 35.29 |
| 'Spinning Spur Wind Ranch' | 161 | 70 | 2300 | 80 | 108 | -102.42 | 35.27 |
| 'Stanton Energy Center' | 120 | 80 | 1500 | 80 | 77 | -101.82 | 32.25 |
| 'Stella' | 201 | 67 | 3000 | 87.5 | 125 | -97.57 | 26.86 |
| 'Stephens Ranch 1' | 211.22 | 118 | 1790 | 80 | 100 | -101.65 | 32.85 |
| 'Stephens Ranch II' | 164.68 | 92 | 1790 | 80 | 100 | -101.73 | 32.97 |
| 'Sweetwater 3' | 135 | 90 | 1500 | 80 | 77 | -100.48 | 32.33 |
| 'Sweetwater IVa' | 135 | 135 | 1000 | 69 | 61.4 | -100.51 | 32.28 |
| 'Sweetwater IVb' | 105.8 | 46 | 2300 | 80 | 93 | -100.47 | 32.29 |
| 'Sweetwater Phase III' | 135 | 90 | 1500 | 80 | 77 | -100.44 | 32.28 |
| 'Tahoka Wind' | 300 | 120 | 2500 | 89 | 127 | -101.65 | 33.12 |
| 'Torrecillas' | 300 | 120 | 2500 | 89 | 127 | -98.79 | 27.62 |
| 'Trent Mesa' | 124.5 | 83 | 1500 | 65 | 82.5 | -100.19 | 32.43 |
| 'Trinity Hills' | 225 | 90 | 2500 | 80 | 96 | -98.68 | 33.44 |
| 'Turkey Track' | 169.5 | 113 | 1500 | 80 | 77 | -100.20 | 32.22 |
| 'Tyler Bluff' | 125.58 | 52 | 2415 | 80 | 108 | -97.40 | 33.72 |
| 'Wake' | 257.25 | 150 | 1715 | 80 | 103 | -101.09 | 33.83 |
| 'Wildcat Ranch' | 150.59 | 67 | 1715 | 80 | 103 | -102.94 | 33.50 |
| 'Wildorado' | 161 | 70 | 2300 | 80 | 93 | -102.30 | 35.28 |
| 'Willow Springs Wind Farm' | 250 | 100 | 2500 | 90 | 116 | -99.69 | 33.39 |
| 'Wolf Ridge' | 112.5 | 75 | 1500 | 80 | 82.5 | -97.38 | 33.73 |

Note that the data in Table A1 is retrieved from the U.S. Wind Turbine Database (USWTDB, https://eerscmap.usgs.gov/uswtdb/).

### Data Availability Statement

Data not available due to [ethical/legal/commercial] restrictions.

### CRediT authorship contribution statement

**Linyue Gao:** Conceptualization, Methodology, Validation, Formal analysis, Investigation, Writing – original draft. **Teja Dasari:** Resources, Writing – review & editing. **Jiarong Hong:** Writing – review & editing, Supervision, Resources.

### Declaration of Competing Interest

The authors declare that they have no known competing financial interests or personal relationships that could have appeared to influence the work reported in this paper.

### Acknowledgments

This work was supported by the National Science Foundation CAREER award (NSF-CBET-1454259), Xcel Energy through the Renewable Development Fund (grant RD4-13) as well as IonE and REU Fellowship of University of Minnesota. The authors would thank Christopher Milliren, Associate Engineer at St. Anthony Falls Laboratory, for the fruitful discussion regarding the Eolos database.